%% file: main.tex
\documentclass[runningheads]{llncs}

\input{preamble}

\begin{document}

\input{frontmatter/llncs}

\input{chapters/introduction}
\input{chapters/background}
\input{chapters/approach}
\input{chapters/evaluation}
\input{chapters/related_work}
\input{chapters/conclusion}

\bibliography{references}

\end{document}

%% file: frontmatter/llncs.tex
\title{Generics-Aware Fuzz Target Generation for Rust Libraries via Structured API Analysis}
\titlerunning{Generics-Aware Fuzz Target Generation for Rust Libraries}


\author{
    Yiming Chen\inst{1}\orcidID{0009-0007-7021-365X} \and
    Kaiwen Zhang\inst{1}\orcidID{0000-0003-4573-8968} \and
    Guanjun Liu\inst{1}\orcidID{0000-0002-7523-4827} \and
    Xiaofeng Li\inst{2} \and
    Xiaogang Dong\inst{2}
}
\authorrunning{Y. Chen et al.}
\institute{%
    School of Computer Science and Technology, Tongji University, Shanghai, China
    \email{\{chenyiming2024,zhangkw,liuguanjun\}@tongji.edu.cn}
    \and
    Beijing Institute of Control Engineering, Beijing, China
    \email{li\_x\_feng@126.com, dongxiaogang@263.net}
}

\bibliographystyle{splncs04}

\maketitle

\begin{abstract}
Fuzzing Rust library APIs requires constructing well-typed, compilable call sequences that satisfy ownership rules, generic parameters, and trait bounds; existing tools ignore these constraints or use shallow heuristics, yielding low coverage. We present \tool{}, which extracts structured API information from Rust documentation, builds an API dependency graph via recursive generics-aware type matching, and uses topology-guided traversal plus LLM synthesis with compiler-error feedback to produce compilable fuzz targets. On 13 crates from \textit{crates.io}, \tool{} achieves 80.75\% macro-average API coverage at 96.19\% compilation success, outperforming RULF and RPG by 4.76$\times$ and 2.43$\times$, and reaching 1.41$\times$ the average API coverage of deepSURF on crates with unsafe-reaching APIs.
\end{abstract}

%% file: chapters/introduction.tex
\section{Introduction}

Rust is increasingly adopted for safety-critical systems due to its compile-time guarantees on memory safety and data race freedom~\cite{IsRustUsedSafely, RustForLinux}, and library fuzzing is widely used to detect bugs that arise despite these guarantees~\cite{Xu2020MemorySafetyCC, UnderstandingMemoryAnd, CloserLookSecurityRisks}.

Rust library fuzzing is challenging, however, because valid fuzz targets must construct API sequences that simultaneously satisfy ownership rules, generic type parameters, and trait bounds. Generating inputs for individual functions in isolation fails to capture inter-API dependencies: effective testing requires sequences of mutually-dependent calls in which each invocation produces type-consistent values consumed by later calls. Existing tools either ignore generic type relationships entirely or rely on shallow heuristics, resulting in low API coverage and generation failures on complex generic crates~\cite{rulf, rpg}.

To address this challenge, we present Generics-aware Rust API Fuzz Target generator (\tool{}). \tool{} employs a three-stage pipeline: (1)~structured API analysis extracts type requirements---including generic parameters, trait bounds, and return types---from Rust documentation; (2)~a generics-aware recursive type matching algorithm determines which API outputs are type-compatible with which parameters, yielding a directed API dependency graph; and (3)~a large language model (LLM)~\cite{Chen2021Codex} translates topology-guided sequences from this graph into compilable fuzz targets, with failed compilations iteratively refined using compiler error feedback.

In summary, this paper makes the following contributions:
\begin{enumerate}[leftmargin=*,label=(\arabic*)]
\item We propose a generics-aware recursive type matching algorithm that explicitly resolves generic parameters, trait bounds, and composite type structures, and tracks constraint satisfaction across API call boundaries to determine whether an API output can serve as a parameter of a given type.
\item Based on this algorithm, we design a directed API dependency graph and a topology-guided API call sequence generation procedure that targets maximum API-level coverage.
\item We implement the proposed approach in \tool{} and evaluate it on 13 real-world Rust libraries sourced from \textit{crates.io}. \tool{} achieves a macro-average API coverage of 80.75\% and a macro-average compilation success rate of 96.19\%. On the eight crates where GRAFT, RULF, and RPG all produce runnable fuzz targets, \tool{} achieves 4.76$\times$ and 2.43$\times$ the average API coverage of RULF and RPG, respectively. On the eight crates with unsafe-reaching APIs, \tool{} outperforms deepSURF---a concurrent LLM-augmented tool targeting memory-safety vulnerability detection---by 1.41$\times$ in average API coverage.
\end{enumerate}

%% file: chapters/background.tex
\section{Background and Motivation}

\subsection{The Rust Programming Language}

Rust is a modern systems programming language designed to provide strong memory safety guarantees without relying on garbage collection. Unlike traditional systems languages such as C and C++, Rust enforces memory safety through a combination of ownership, borrowing, and lifetime rules that are statically checked at compile time. These mechanisms regulate aliasing, mutation, and resource lifetimes in a disciplined manner, significantly reducing the likelihood of common memory safety errors~\cite{rustbelt, stacked-borrows}.

To support low-level systems programming and interoperability, Rust provides an \texttt{unsafe} mechanism that allows developers to bypass certain compile-time checks. Code marked as \texttt{unsafe} may perform operations such as raw pointer manipulation, manual memory management, or foreign function interface calls, which are beyond the guarantees enforced by the Rust type system. Although \texttt{unsafe} code is intended to be used sparingly and encapsulated behind safe abstractions, incorrect usage can still lead to memory-related vulnerabilities in practice~\cite{IsRustUsedSafely, Xu2020MemorySafetyCC}. As a result, memory-related bugs can still arise in real-world Rust software.

In addition to its ownership model, Rust features a highly expressive type system underlying both its safety guarantees and zero-cost abstractions. Rust extensively relies on generics to enable reusable and zero-cost abstractions. These generics include type generics, lifetime generics, and constant generics, often combined with trait bounds to constrain permissible instantiations. Such rich type constructs enable powerful abstractions but also introduce significant complexity for program analysis and automated test generation~\cite{rulf, rpg}.

\subsection{Motivating Example}

\begin{lstlisting}[caption={A generic Rust example with trait bounds and unsafe code}, label={lst:motivating}]
pub trait Element {}
impl Element for u8 {}
impl Element for i32 {}

pub struct Buffer<T: Element> {
    ptr: NonNull<T>,
    len: usize,
}

impl<T: Element> Buffer<T> {
    pub fn new(slice: &mut [T]) -> Self {
        Self {
            ptr: NonNull::new(slice.as_mut_ptr()).unwrap(),
            len: slice.len(),
        }
    }

    pub unsafe fn get_unchecked(&self, index: usize) -> &T {
        &*self.ptr.as_ptr().add(index)
    }
}

pub fn read_element<T: Element>(buf: &Buffer<T>, index: usize) -> &T {
    unsafe { buf.get_unchecked(index) }
}
\end{lstlisting}

\Cref{lst:motivating} shows a generic buffer \texttt{Buffer<T>} with bound \texttt{T: Element} and an \texttt{unsafe} method \texttt{get\_unchecked}.
Fuzzing \texttt{read\_element<T>} requires a \texttt{\&Buffer<T>} from \texttt{Buffer::new} (itself needing \texttt{\&mut [T]} with \texttt{T: Element}): \texttt{T} and the trait bound must stay consistent across calls, so no API can be fuzzed in isolation.
Shallow generators that ignore these multi-call dependencies cannot build such sequences.

\begin{figure}[t]
\centering
\includegraphics[width=\linewidth,height=0.20\textheight,keepaspectratio]{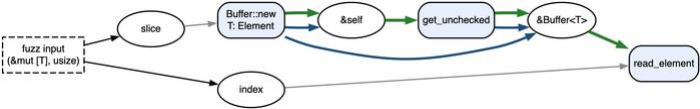}
\caption{API dependency graph generated by \tool{} for \Cref{lst:motivating}, annotated with additional information.}
\label{fig:api-graph}
\end{figure}

As illustrated in \Cref{fig:api-graph}, \tool{} addresses this gap by constructing a generics-aware API dependency graph.
The blue edges are generics-aware type-flow edges that remain undiscovered under purely syntactic matching.
Because \tool{} prefers deeper nodes during traversal, it recovers the green call chain
$\texttt{Buffer::new}\!\rightarrow\!\texttt{\&self}\!\rightarrow\!\texttt{get\_unchecked}\!\rightarrow\!\texttt{\&Buffer<T>}\!\rightarrow\!\texttt{read\_element}$,
exercising all three APIs.
Without generics-aware matching, no typed producer of \texttt{\&Buffer<T>} is found and \texttt{read\_element} remains unreachable.

%% file: chapters/approach.tex
\section{Approach}

\subsection{Overall Framework}

\begin{figure}[t]
    \centering
    \includegraphics[width=\linewidth,height=0.22\textheight,keepaspectratio]{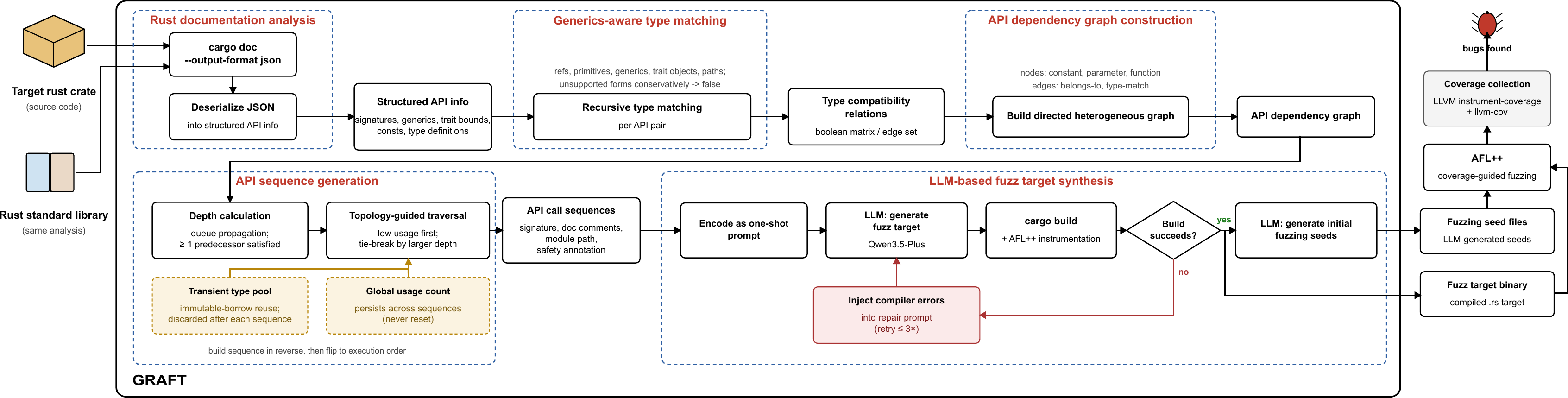}
    \caption{Overview of \tool{}}
    \label{fig:overview}
\end{figure}

\tool{} is a generics-aware fuzz target generation framework for Rust libraries that automatically produces compilable fuzz targets and initial seeds for use with coverage-guided fuzzers such as AFL++~\cite{aflplusplus}. As illustrated in \Cref{fig:overview}, the framework consists of five stages.

\begin{itemize}

  \item \textit{Rust Documentation Analysis.}
  \tool{} invokes \texttt{cargo doc} to obtain a structured, compiler-produced view of the crate’s public API surface. It extracts function signatures, generic parameters, trait bounds, public constants, and type definitions, preserving all information in structured form to support precise reasoning in later stages.

  \item \textit{Generics-aware Type Matching.}
  For each pair of APIs, \tool{} determines whether the output of one API can satisfy a parameter of another via a recursive type matching algorithm. The algorithm handles references, primitives, generic parameters, trait objects, and resolved path types, and conservatively returns $\bot$ for unsupported forms to avoid introducing infeasible dependencies.

  \item \textit{API Dependency Graph Construction.}
  Based on the type matching results, \tool{} builds a directed heterogeneous graph whose nodes represent public constants, function parameters, and functions, and whose edges encode parameter ownership and type-compatible value-flow relationships among APIs.

  \item \textit{API Sequence Generation.}
  Guided by the dependency graph, \tool{} generates type-consistent API call sequences using a topology-guided traversal strategy. APIs are prioritized first by usage count and then by dependency depth to maximize API-level coverage, and a transient type pool enables value reuse for immutable-borrow parameters.

  \item \textit{LLM-based Fuzz Target Synthesis.}
  Each API sequence is translated into a self-contained, compilable Rust fuzz target by a large language model~\cite{Chen2021Codex}. The LLM resolves type instantiations, ownership patterns, and necessary binding code, and is retried on compilation failure with compiler error feedback. Initial fuzzing seeds are generated alongside each target.

\end{itemize}

\subsection{Rust Documentation Analysis}

\tool{} builds its analysis on top of Rust’s official documentation infrastructure.
Given a target crate, \tool{} invokes \texttt{cargo doc} with JSON output enabled and deserializes the result into a structured \texttt{rustdoc\_types::Crate} object, providing a compiler-produced view of the crate’s public API surface.
For each public function or method, \tool{} records its parameter list, return type, generic parameters, and declared trait bounds; public types are collected with their generic parameters and implemented traits; and public constants are recorded as potential value sources.

The same pipeline is applied to the Rust standard library for commonly used standard types.
All type information is preserved in its structured \texttt{rustdoc\_types} form rather than reduced to textual signatures, enabling precise reasoning about nested generic instantiations in later stages.

\subsection{Generics-aware Type Matching}
\label{subsec:type-matching}

Determining whether one API's output can feed another is non-trivial in Rust because generics and trait bounds are pervasive. \tool{} addresses this with a generics-aware recursive type matching algorithm.

\begin{definition}[Type Compatibility]
\label{def:type-compat}
Let $\mathit{from}$ and $\mathit{to}$ be Rust types drawn from the public API surface of a crate.
$\mathit{from}$ is \emph{type-compatible} with $\mathit{to}$, written $\textsc{Match}(\mathit{from}, \mathit{to}) = \top$, if a value of type $\mathit{from}$ can serve as an argument for a parameter of type $\mathit{to}$.
$\textsc{Match}$ is a conservative approximation: it returns $\bot$ whenever compatibility cannot be established with confidence, to avoid introducing infeasible API dependencies.
\end{definition}

For each source--target API pair, a checker is parameterized by both APIs' generic parameters and trait bounds (including those inherited from enclosing \texttt{struct}/\texttt{enum}/\texttt{impl} blocks) and a global type pool of definitions and implemented traits.
It recurses on $\mathit{to}$: wrappers (references, slices, arrays, pointers) are stripped; generic/\texttt{impl}/\texttt{dyn} targets require all trait bounds of $\mathit{from}$ to hold; resolved-path types use structural equality; unsupported forms yield $\bot$.
\Cref{tab:type-matching-summary} summarizes these rules.

\begin{table}[t]
\centering
\caption{Summary of generics-aware type matching rules.}
\label{tab:type-matching-summary}
\scriptsize
\setlength{\tabcolsep}{2.5pt}
\begin{tabular}{@{}lll@{}}
\toprule
\makecell[l]{Target Type\\(\textit{to})} & \makecell[l]{Source Type\\(\textit{from})} & \makecell[l]{Matching\\Strategy} \\
\midrule
\makecell[l]{Reference / Slice /\\Array / Pointer} & Any & \makecell[l]{Recursively match\\inner type} \\
Primitive parameter & \makecell[l]{External fuzz\\input} & \makecell[l]{Satisfied directly;\\no producer edge} \\
\makecell[l]{Generic / \texttt{impl} /\\\texttt{dyn} trait} & Any & \makecell[l]{Check required\\trait bounds} \\
Resolved path & Resolved path & Structural equality \\
\makecell[l]{Tuple (single\\non-primitive)} & Any & \makecell[l]{Match non-primitive\\element} \\
Unsupported forms & Any & Conservatively fail \\
\bottomrule
\end{tabular}
\end{table}

Trait-bound checks query the type pool (std impls for primitives, documented impls for resolved paths, recursive handling for composites) and succeed only when every required bound of $\mathit{to}$ is matched by name and recursively compatible type arguments.
The implementation covers inline bounds and simple \texttt{where T: Trait} predicates, but not associated-type predicates, trait aliases, negative/auto-trait reasoning, or higher-ranked lifetimes; such cases fail conservatively (possible false negatives).

\subsection{API Dependency Graph Construction}

\begin{definition}[API Dependency Graph]
\label{def:api-graph}
The \emph{API dependency graph} is a directed heterogeneous graph $G = (V, E)$ over a crate's public API surface.
The node set $V = V_c \cup V_p \cup V_f$ comprises public constant nodes $V_c$, function parameter nodes $V_p$, and function nodes $V_f$ (methods are treated uniformly as functions; type definitions are not graph nodes).
The edge set $E = E_{\mathrm{own}} \cup E_{\mathrm{flow}}$ comprises:
\begin{itemize}[noitemsep,topsep=2pt]
  \item \emph{parameter ownership edges} $E_{\mathrm{own}} \subseteq V_p \times V_f$, connecting each parameter node to its owning function; and
  \item \emph{type-flow edges} $E_{\mathrm{flow}} \subseteq (V_c \cup V_f) \times V_p$, where $(v, p) \in E_{\mathrm{flow}}$ iff $\textsc{Match}(\tau(v), \tau(p)) = \top$ (Definition~\ref{def:type-compat}), with $\tau(\cdot)$ denoting the associated type: the return type for $v \in V_f$, the parameter type for $p \in V_p$, and the declared type for $v \in V_c$. In the implementation, $\tau(v)$ for $v \in V_f$ denotes the unwrapped return type: \texttt{Result<T,~E>} and \texttt{Option<T>} are expanded to their inner type $T$; tuple and \texttt{Box} wrappers are handled as composite types per Definition~\ref{def:type-compat}.
\end{itemize}
\end{definition}

A function node $f \in V_f$ is \emph{reachable} in $G$ if
\[
\begin{aligned}
  &\forall\, p \in \bigl\{p' \in V_p \mid (p', f) \in E_{\mathrm{own}}\bigr\},\\
  &\quad \mathrm{prim}(\tau(p)) \vee \exists\, v \in \bigl(V_c \cup (V_f \setminus \{f\})\bigr) : (v, p) \in E_{\mathrm{flow}},
\end{aligned}
\]
where $\mathrm{prim}(\tau(p))$ holds when $\tau(p)$ is a primitive type whose values are directly supplied by the fuzzer without a producer API, and $f$ is excluded from the producer set to avoid circular self-dependency.
Such edges represent type-compatible candidate value flows; they do not by themselves guarantee that borrowing, ownership transfer, or moves are realizable in Rust code.
\Cref{fig:api-graph} instantiates this construction on the motivating \texttt{Buffer<T>} example: \texttt{Buffer::new} feeds parameters requiring \texttt{\&Buffer<T>} with consistent $T$, while per-sequence usage bounds control cycles.

\subsection{API Sequence Generation}
\label{subsec:api-sequence}

\begin{definition}[API-Level Coverage]
\label{def:coverage}
Given $G = (V, E)$ (Definition~\ref{def:api-graph}), a sequence $s = (f_1, \ldots, f_k) \in V_f^+$ is \emph{type-consistent} if for every $f_i$ and each parameter $p$ of $f_i$:
\[
\begin{aligned}
  &\mathrm{prim}(\tau(p))\ \vee\ \exists\,c \in V_c: (c,p) \in E_{\mathrm{flow}}\\
  &\quad\vee\ \exists\,j < i: \textsc{Match}(\tau(f_j), \tau(p)) = \top,
\end{aligned}
\]
where $\textsc{Match}$ is as in Definition~\ref{def:type-compat}.
Let $\mathcal{S}$ be a set of type-consistent sequences. The \emph{API-level coverage} of $\mathcal{S}$ is:
\[
  \mathrm{Cov}(\mathcal{S}) = \frac{|\{f \in V_f : \exists\, s \in \mathcal{S},\, f \in s\}|}{|V_f|}
\]
\end{definition}

\tool{} maximizes API-level coverage by generating type-consistent sequences via topology-guided traversal.
Each reachable function gets a \emph{depth}: $d(v)=d(f)=0$ for constants and all-primitive APIs; otherwise
\[
  d(f) = \max_{\substack{p:\,(p,f)\in E_{\mathrm{own}} \\ \neg\,\mathrm{prim}(\tau(p))}}\ \min_{v:\,(v,p)\in E_{\mathrm{flow}}}\ \bigl(d(v) + 1\bigr).
\]
Unreachable nodes have undefined depth and are excluded from target selection.
Depth is computed by queue propagation (a parameter needs any predecessor; an API needs all parameters), rather than a classical topological sort.

\Cref{alg:api-sequence} constructs one sequence.
Targets are chosen by low global usage, then high depth; a per-sequence bound $B$ guarantees termination.
Construction proceeds backward from the target: primitives are skipped, immutable-borrow needs may reuse a transient type pool, and other parameters enqueue a least-used producer under $B$.
The collected list is reversed so producers precede consumers.
An outer driver repeats the procedure until a coverage quota or a requested sequence count is met.
Each sequence carries signatures, docs, module paths, and safety qualifiers for LLM synthesis.

\begin{algorithm}[htbp]
\caption{Heuristic API Sequence Generation with Type Pool}
\label{alg:api-sequence}
\KwIn{API dependency graph $G$ with API depths, maximum per-sequence usage bound $B$}
\KwOut{Candidate API call sequence $S$}

\eIf{a target API is specified}{
  $a \leftarrow$ specified target API\;
}{
  $a \leftarrow$ API with minimum usage count and maximum depth\;
}

Initialize sequence $S \leftarrow [a]$\;
Initialize empty type pool $P$\;
Initialize queue $Q \leftarrow [a]$\;
Initialize per-sequence usage counter $C[x] \leftarrow 0$ for each API $x$\;
Increment global usage count of $a$ and set $C[a] \leftarrow 1$\;

\While{$Q$ is not empty}{
  Pop an API $u$ from $Q$\;
  \ForEach{parameter $p$ of $u$}{
    \uIf{$p$ is primitive-typed}{
      continue\;
    }
    \uElseIf{$p$ requires only an immutable reference and a compatible producer exists in $P$}{
      continue\;
    }
    Select a value-producing API $b$ that can satisfy $p$, with minimum global usage count and $C[b] < B$\;
    \If{such a producer exists}{
      Add $b$ to $S$ and $Q$\;
      Increment global usage count of $b$ and set $C[b] \leftarrow C[b] + 1$\;
      \If{$p$ requires only an immutable reference}{
        Add $b$ to $P$\;
      }
    }
    \Else{
      Mark $p$ unresolved in the current pass and continue\;
    }
  }
}

Reverse $S$\;
\Return $S$\;
\end{algorithm}

On \Cref{fig:api-graph}, the green edges depict the recovered chain
$\texttt{Buffer::new}\!\rightarrow\!\texttt{\&self}\!\rightarrow\!\texttt{get\_unchecked}\!\rightarrow\!\texttt{\&Buffer<T>}\!\rightarrow\!\texttt{read\_element}$.

\subsection{LLM-based Fuzz Target Synthesis}
\label{subsec:llm-synthesis}

\tool{} uses an LLM~\cite{Chen2021Codex} only to concretize sequences into compilable fuzz targets and seeds; sequence choice and coverage decisions remain graph-driven.
Each sequence is encoded as a one-shot JSON prompt (signatures with generics/bounds, docs, module paths; no bodies).
Three templates cover generation, repair (compiler errors after failed builds), and seed input synthesis; exact prompts are in the artifact.
Experiments use \texttt{qwen3.5-plus} via DashScope (temperature $0.0$, 8192-token cap).
The model emits a self-contained AFL++-ready program~\cite{aflplusplus}; on failure, repair retries up to three times.
Seeds are written as a JSON-derived corpus into the fuzzer input directory.

%% file: chapters/evaluation.tex
\section{Evaluation}

\subsection{Evaluation Setup}

\paragraph{Environment.}
All experiments are conducted on a server equipped with two Intel Xeon Silver 4314 processors (16 cores / 32 threads each, 2.40\,GHz base, 3.40\,GHz boost) and 128\,GB RAM, running Rocky Linux 9.7. The implementation uses Rust stable 1.94.1 and nightly 1.97.0-nightly, AFL++ 4.40c through \texttt{afl.rs}, and clang/LLVM 20.1.8. \tool{} uses \texttt{qwen3.5-plus} through the DashScope OpenAI-compatible API for fuzz target synthesis. The artifact package includes raw command lines for all pipeline stages.

\paragraph{Benchmark Crates.}
We evaluate \tool{} on 13 widely-used Rust libraries sourced from \textit{crates.io}, covering diverse application domains including encoding (\texttt{base64}, \texttt{byteorder}), Unicode processing (\texttt{unicode-segmentation}, \texttt{idna}), data structures (\texttt{fixedbitset}, \texttt{bumpalo}), serialization (\texttt{csv}), date and time handling (\texttt{chrono}, \texttt{time}), byte manipulation (\texttt{bytes}), and text processing (\texttt{regex}, \texttt{ryu}, \texttt{autocfg}). These crates span a wide range of API set sizes (5 to 518 public APIs) and complexity levels, and include crates on which both RULF and RPG are known to fail or underperform. The evaluated versions are \texttt{base64} 0.22.1, \texttt{bumpalo} 3.20.2, \texttt{byteorder} 1.5.0, \texttt{autocfg} 1.5.0, \texttt{fixedbitset} 0.5.7, \texttt{idna} 1.1.0, \texttt{regex} 1.12.3, \texttt{ryu} 1.0.23, \texttt{time} 0.3.47, \texttt{unicode-segmentation} 1.13.2, \texttt{chrono} 0.4.44, \texttt{bytes} 1.11.1, and \texttt{csv} 1.4.0. All crates are evaluated with their default feature sets.

\paragraph{Baselines.}
We compare \tool{} against three baselines. RULF~\cite{rulf} generates fuzz targets via API dependency graph traversal (up to 10 targets per crate). RPG~\cite{rpg} employs a pool-based BFS strategy with generic support (up to 300 targets per crate). deepSURF~\cite{deepsurf} is a concurrent LLM-augmented Rust fuzzing tool that uses a custom compiler plugin (\texttt{rustc\_surf}) to identify unsafe-reaching APIs (URAPIs) and generates fuzz targets targeting those APIs; it is designed for memory-safety vulnerability detection rather than broad API coverage. For the deepSURF comparison, we use the same crate versions, LLM backend (\texttt{qwen3.5-plus} via DashScope), coverage measurement methodology, and fuzzing budget (3 hours per fuzz target with AFL++) as \tool{}.

\paragraph{Metrics.}
We measure two primary metrics: (1) \textit{compilation success rate}, defined as the proportion of generated fuzz targets that compile successfully, and (2) \textit{API coverage rate}, defined as the proportion of public APIs in a crate that are exercised by at least one compiled fuzz target. The API denominator is derived from the rustdoc-extracted public API surface after two filtering steps. \texttt{filter\_api()} removes 28 universally-implemented methods by name---type conversions (\texttt{into}, \texttt{from}, \texttt{try\_into}, \texttt{try\_from}), operator overloads (\texttt{add}, \texttt{bit\_and}, etc.), iterator adapters (\texttt{iter}, \texttt{into\_iter}), and serde helpers (\texttt{serialize}, \texttt{deserialize})---as these reflect uniform language-level behavior rather than library-specific functionality; type conversions among these are handled implicitly by the LLM during synthesis. \texttt{filter\_types()} removes enums that represent error types, as these serve as return types rather than callable API entry points. Across all 13 benchmark crates, filtering reduces the API count from 4,258 to 1,745 (59.02\% removed). To confirm that this does not inflate coverage numbers, \tool{}'s dependency graph connects 3,853 of the 4,258 unfiltered APIs (90.49\% graph reachability) even without filtering; actual post-synthesis coverage on the unfiltered surface would be slightly lower due to LLM compilation failures, but the graph structure demonstrates that the approach is effective across the full unfiltered API surface. Re-exported public items, trait-implementation methods, and macro-generated public APIs are counted whenever rustdoc exposes them in the crate's public interface. Coverage is measured using LLVM's \texttt{instrument-coverage} instrumentation together with \texttt{llvm-cov export}. Each compiled fuzz target is fuzzed with AFL++~\cite{aflplusplus} for 3 hours; post-fuzzing API coverage is computed by replaying the accumulated corpus through LLVM-instrumented binaries, following the same methodology as the baselines.

\tool{} targets the \emph{fuzz target generation} problem: given a Rust library, automatically produce compilable, type-correct fuzz targets with broad API coverage. Consistent with RULF~\cite{rulf} and RPG~\cite{rpg}---the closest prior tools---we evaluate on compilation success rate and API coverage as direct measures of target quality; bug-finding effectiveness is a downstream property of coverage breadth, discussed further in the threats to validity section.

The experiments address the following research questions:

\begin{enumerate}[label=\textbf{RQ\arabic*.}]
    \item How effective is \tool{} in generating compilable fuzz targets with high API coverage?
    \item What is the contribution of each component of \tool{} to fuzz target quality?
    \item How does \tool{} compare to state-of-the-art approaches RULF~\cite{rulf}, RPG~\cite{rpg}, and deepSURF~\cite{deepsurf} in terms of API coverage?
\end{enumerate}

\subsection{RQ1: Quality of Generated Fuzz Targets}

\Cref{tab:rq1} presents the compilation success rates and API coverage rates achieved by \tool{} across all 13 benchmark crates. Overall, \tool{} generates 1467 fuzz targets in total, of which 1421 compile successfully, corresponding to a micro-average compilation success rate of 96.86\%. At the crate level, the macro-average compilation success rate is 96.19\%. API coverage ranges from 62.07\% (\texttt{byteorder}) to 100\% (\texttt{ryu} and \texttt{bumpalo}), with a macro-average of 80.75\% across all 13 crates.

\begin{table}[t]
\centering
\caption{Fuzz target quality of \tool{} on 13 benchmark crates.}
\label{tab:rq1}
\scriptsize
\setlength{\tabcolsep}{2.5pt}
\begin{tabular}{lcccc c}
\toprule
\textbf{Crate} & \textbf{Targets} & \textbf{Compiled} & \textbf{Comp.\%} & \textbf{Covered} & \textbf{API Cov.\%} \\
\midrule
ryu                  & 6   & 6   & 100.00\% & 5/5     & 100.00\% \\
bumpalo              & 24  & 24  & 100.00\% & 44/44   & 100.00\% \\
base64               & 23  & 23  & 100.00\% & 24/26   &  92.31\% \\
regex                & 150 & 142 &  94.67\% & 183/210 &  87.14\% \\
idna                 & 10  & 10  & 100.00\% & 15/18   &  83.33\% \\
fixedbitset          & 60  & 57  &  95.00\% & 75/91   &  82.42\% \\
chrono               & 534 & 524 &  98.13\% & 404/518 &  77.99\% \\
time                 & 381 & 375 &  98.43\% & 316/409 &  77.26\% \\
csv                  & 43  & 35  &  81.40\% & 113/148 &  76.35\% \\
autocfg              & 51  & 51  & 100.00\% & 23/31   &  74.19\% \\
bytes                & 109 & 105  &  96.33\% & 108/152 &  71.05\% \\
unicode-segmentation & 24  & 24  & 100.00\% & 23/35   &  65.71\% \\
byteorder            & 52  & 45  &  86.54\% & 36/58   &  62.07\% \\
\midrule
\textbf{Macro average} & --  & --  & \textbf{96.19\%} & -- & \textbf{80.75\%} \\
\bottomrule
\end{tabular}
\end{table}

The compilation success rate exceeds 94\% for 10 of the 13 crates; the two outliers---\texttt{csv} (81.40\%) and \texttt{byteorder} (86.54\%)---involve complex generic trait bounds and type conversions that occasionally require more than the allotted retry budget to resolve.

In terms of API coverage, \tool{} achieves 100\% coverage for \texttt{ryu} and \texttt{bumpalo} and above 90\% for \texttt{base64} and \texttt{regex}. Even for the largest crate in the benchmark (\texttt{chrono}, 518 public APIs), \tool{} covers 77.99\% of APIs. The generics-aware type matching algorithm and dependency-guided traversal strategy enable \tool{} to reach APIs that are only accessible through specific sequences of type-constrained calls.

\subsection{RQ2: Ablation Study}

To assess the individual contribution of each structural component, we evaluate three degraded variants of \tool{} on five representative crates (\texttt{base64}, \texttt{byteorder}, \texttt{fixedbitset}, \texttt{time}, \texttt{regex}), generating 20 fuzz targets per crate and fuzzing each for one hour with AFL++~\cite{aflplusplus}.

\begin{itemize}
  \item \textbf{w/o generic matching}: The generics-aware type matching is replaced by syntactic equality; generic parameters, trait objects, and \texttt{impl Trait} bounds are not resolved.
  \item \textbf{random sequence assembly}: The dependency graph is still constructed, but API sequences are assembled by random sampling rather than graph-guided traversal, so type compatibility between consecutive calls is not guaranteed.
  \item \textbf{w/o topo traversal}: The dependency graph is used, but the topology-guided target selection (prioritizing APIs by dependency depth and usage count) is replaced by random selection.
\end{itemize}

\paragraph{Impact on dependency graph structure.}
Disabling generics-aware matching drastically reduces dependency graph edges: across the four crates with non-zero edges, the reduction averages 84.5\%, ranging from 56.7\% (\texttt{time}) to 99.97\% (\texttt{regex}). The near-total collapse for \texttt{regex} reflects that almost all of its APIs rely on generic or trait-bounded types, leaving almost no type-compatible connections identifiable without generics-aware reasoning.

\begin{figure}[t]
  \centering
  \includegraphics[width=\linewidth,height=0.20\textheight,keepaspectratio]{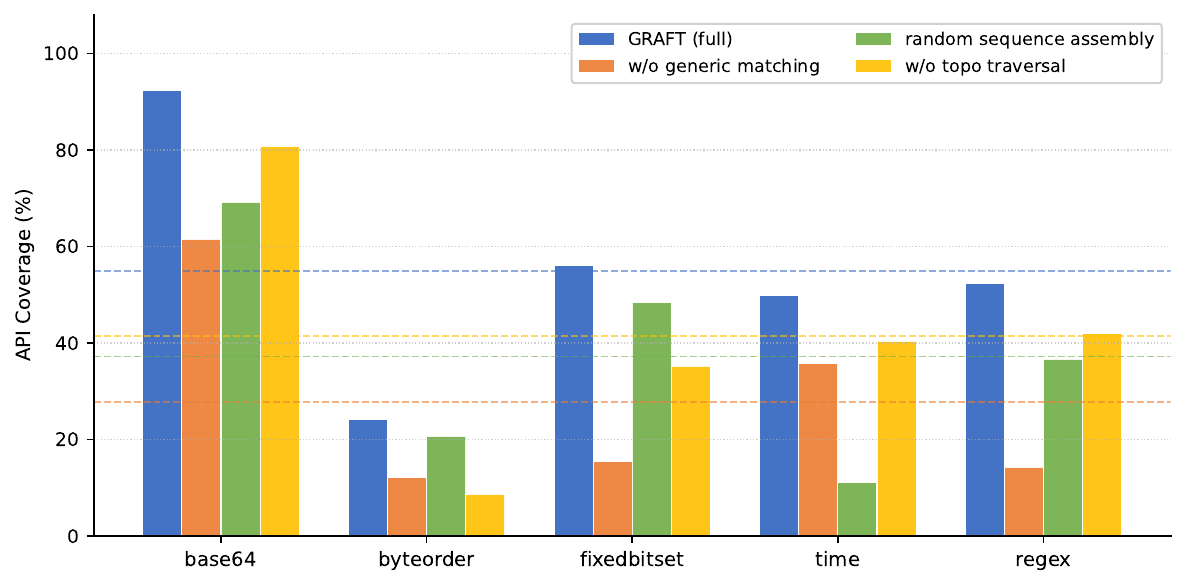}
  \caption{API coverage of \tool{} and ablation variants across five benchmark crates. Dashed lines indicate per-variant averages.}
  \label{fig:rq2}
\end{figure}

\paragraph{Impact on API coverage.}
\Cref{tab:ablation-cov} and \Cref{fig:rq2} compare the API coverage of the full \tool{} against each ablation variant. Removing any single component substantially degrades coverage. The full \tool{} achieves an average of 54.95\% across the five crates, compared to 41.36\% without topology-guided traversal, 37.19\% with random sequence assembly, and 27.80\% without generics-aware type matching.

\begin{table}[t]
\centering
\caption{API coverage (\%) of \tool{} and ablation variants.}
\label{tab:ablation-cov}
\scriptsize
\setlength{\tabcolsep}{2.5pt}
\begin{tabular}{@{}lcccc@{}}
\toprule
\textbf{Crate} & \textbf{\tool{}} & \makecell[c]{\textbf{w/o}\\\textbf{generic}} & \makecell[c]{\textbf{random seq.}\\\textbf{assembly}} & \makecell[c]{\textbf{w/o}\\\textbf{topo}} \\
\midrule
base64      & \textbf{92.31} & 61.54 & 69.23 & 80.77 \\
byteorder   & \textbf{24.14} & 12.07 & 20.69 &  8.62 \\
fixedbitset & \textbf{56.04} & 15.38 & 48.35 & 35.16 \\
time        & \textbf{49.88} & 35.70 & 11.00 & 40.34 \\
regex       & \textbf{52.38} & 14.29 & 36.67 & 41.90 \\
\midrule
\textbf{Average} & \textbf{54.95} & 27.80 & 37.19 & 41.36 \\
\bottomrule
\end{tabular}
\end{table}

The degradation patterns differ by crate characteristics. For \texttt{regex} and \texttt{fixedbitset}, which have high generic density, disabling generics-aware matching is the most damaging: \texttt{regex} drops from 52.38\% to 14.29\% and \texttt{fixedbitset} from 56.04\% to 15.38\%, consistent with the near-total collapse of graph edges observed for these crates. For \texttt{time}, whose APIs form strong construction chains, removing the dependency graph causes the sharpest drop (49.88\% $\to$ 11.00\%), confirming that random sequence assembly cannot satisfy the type-level ordering required to reach deep API paths. Topology-guided traversal provides a further consistent lift: the w/o-topo variant consistently underperforms full \tool{} across all five crates, indicating that prioritizing under-covered, deeper APIs materially improves coverage breadth. The three components thus contribute independently: their combined effect yields a 27.15 percentage-point improvement in average API coverage over the weakest ablation variant.

\subsection{RQ3: Comparison with State-of-the-Art}

\begin{figure}[t]
  \centering
  \includegraphics[width=\linewidth,height=0.22\textheight,keepaspectratio]{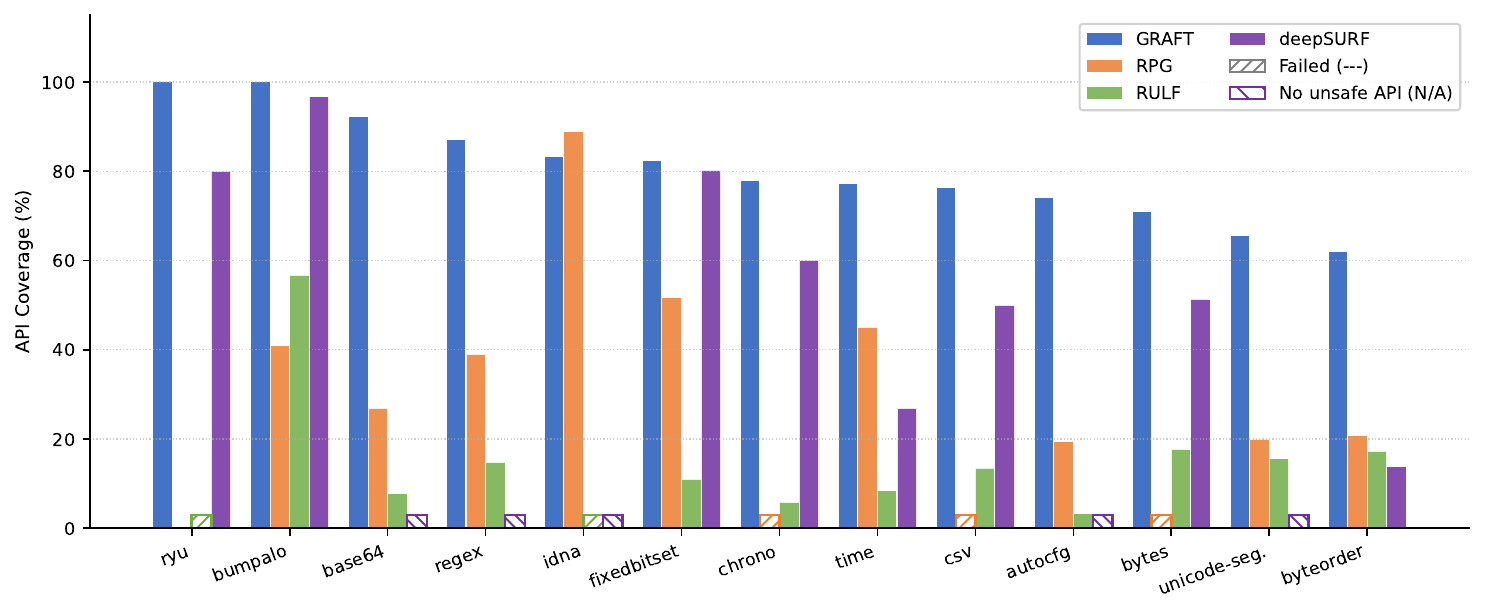}
  \caption{API coverage of \tool{}, RPG, RULF, and deepSURF across 13 benchmark crates. Hatched bars: ``---'' = tool failed; ``N/A'' = no unsafe-reaching APIs (deepSURF skips by design).}
  \label{fig:rq3}
\end{figure}

\Cref{tab:rq3} and \Cref{fig:rq3} compare the API coverage achieved by \tool{}, RPG~\cite{rpg}, RULF~\cite{rulf}, and deepSURF~\cite{deepsurf} across all 13 benchmark crates. RPG fails to produce runnable targets for \texttt{chrono}, \texttt{csv}, and \texttt{bytes}; RULF fails on \texttt{idna} and \texttt{ryu}. deepSURF skips five crates---\texttt{base64}, \texttt{regex}, \texttt{idna}, \texttt{autocfg}, and \texttt{unicode-segmentation}---because its URAPI static analysis detects no unsafe-reaching APIs in them, reflecting deepSURF's design focus on memory-safety vulnerability detection rather than broad API coverage.

\begin{table}[t]
\centering
\caption{API coverage (\%) of \tool{}, RPG, RULF, and deepSURF across 13 benchmark crates.}
\label{tab:rq3}
\scriptsize
\setlength{\tabcolsep}{2.5pt}
\begin{tabular*}{\linewidth}{@{\extracolsep{\fill}}lcccc@{}}
\toprule
\textbf{Crate} & \textbf{\tool{}} & \textbf{RPG} & \textbf{RULF} & \textbf{deepSURF} \\
\midrule
ryu                  & \textbf{100.00\%} &  0.00\%   & ---       &  80.00\%  \\
bumpalo              & \textbf{100.00\%} & 40.91\%   & 56.67\%   &  96.67\%  \\
base64               & \textbf{ 92.31\%} & 26.92\%   &  7.69\%   &  N/A      \\
regex                & \textbf{ 87.14\%} & 39.05\%   & 14.76\%   &  N/A      \\
idna                 &          83.33\%  & \textbf{88.89\%} & ---  &  N/A      \\
fixedbitset          & \textbf{ 82.42\%} & 51.65\%   & 10.99\%   &  80.22\%  \\
chrono               & \textbf{ 77.99\%} &  ---      &  5.79\%   &  60.04\%  \\
time                 & \textbf{ 77.26\%} & 44.99\%   &  8.41\%   &  26.86\%  \\
csv                  & \textbf{ 76.35\%} &  ---      & 13.51\%   &  50.00\%  \\
autocfg              & \textbf{ 74.19\%} & 19.35\%   &  3.23\%   &  N/A      \\
bytes                & \textbf{ 71.05\%} &  ---      & 17.76\%   &  51.32\%  \\
unicode-seg.         & \textbf{ 65.71\%} & 20.00\%   & 15.62\%   &  N/A      \\
byteorder            & \textbf{ 62.07\%} & 20.69\%   & 17.24\%   &  13.79\%  \\
\midrule
\textbf{Avg.~(a)} & \textbf{80.14\%} & 32.95\% & 16.83\% & ---     \\
\textbf{Avg.~(b)} & \textbf{80.89\%} & ---     & ---     & \textbf{57.36\%} \\
\bottomrule
\end{tabular*}

\vspace{2pt}
{\footnotesize\itshape Note: ``---'' = no runnable fuzz targets generated; ``N/A'' = no unsafe-reaching APIs (deepSURF skips by design). Avg.~(a): 8 crates where \tool{}, RPG, and RULF all succeed. Avg.~(b): 8 crates with unsafe-reaching APIs where deepSURF runs.\par}
\end{table}

\paragraph{Comparison with RULF and RPG.}
On the 8 crates where GRAFT, RPG, and RULF all produce runnable results, \tool{} achieves an average API coverage of 80.14\%, compared to 16.83\% for RULF and 32.95\% for RPG---representing 4.76$\times$ and 2.43$\times$ the respective averages. The advantage is particularly pronounced on crates with complex generic APIs: for \texttt{regex}, \tool{} covers 87.14\% of APIs versus 39.05\% (RPG) and 14.76\% (RULF); for \texttt{fixedbitset}, 82.42\% versus 51.65\% (RPG) and 10.99\% (RULF). \tool{} is the only tool to generate runnable fuzz targets for all 13 crates; RPG fails on \texttt{chrono}, \texttt{csv}, and \texttt{bytes} due to compilation errors, while RULF fails on \texttt{idna} and \texttt{ryu} due to syntactic errors and private API access. The single crate where RPG marginally outperforms \tool{} is \texttt{idna} (88.89\% vs.\ 83.33\%), likely owing to \texttt{idna}'s small API surface (18 APIs) and RPG's BFS strategy achieving high coverage with simple fuzz targets.

\paragraph{Comparison with deepSURF.}
On the 8 crates with unsafe-reaching APIs where deepSURF produces results, \tool{} achieves an average API coverage of 80.89\% versus 57.36\% for deepSURF---a 1.41$\times$ improvement. The gap is largest on \texttt{time} (77.26\% vs.\ 26.86\%) and \texttt{byteorder} (62.07\% vs.\ 13.79\%), where deepSURF's fuzz targets are constrained to APIs reachable from unsafe blocks, leaving many safe-but-important APIs uncovered. On crates with dense unsafe usage such as \texttt{bumpalo} (96.67\%) and \texttt{fixedbitset} (80.22\%), deepSURF performs competitively. On the five crates without unsafe-reaching APIs---\texttt{base64}, \texttt{regex}, \texttt{idna}, \texttt{autocfg}, and \texttt{unicode-segmentation}---deepSURF produces no fuzz targets at all, reflecting its fundamental design trade-off between unsafe-API reachability and general API coverage. 

\subsection{Threats to Validity}

\paragraph{Benchmark selection.}
Our evaluation uses 13 widely-used crates from \textit{crates.io}, spanning multiple domains and API sizes (RQ1--RQ3), but they may not represent the full diversity of Rust libraries in terms of macro-heavy codebases, specialized domains, or complex feature configurations. To mitigate this, we deliberately included crates on which prior tools are known to fail or underperform, ensuring the benchmark is sufficiently challenging.

\paragraph{Baseline fairness.}
RULF and RPG are evaluated under their default configurations (RQ3); their target-generation budgets and failure modes differ from \tool{}'s, so absolute coverage comparisons should be interpreted under each tool's reported configuration. Observed failures may stem from unsupported Rust features, synthesis limitations, or budget constraints that are difficult to disentangle without tool-internal instrumentation. To mitigate this, the deepSURF comparison (RQ3) uses identical crate versions, LLM backend, fuzzing budget, and coverage measurement methodology across both tools.

\paragraph{LLM nondeterminism and API counting.}
\tool{} relies on an LLM for fuzz target synthesis (RQ1, RQ2); model outputs may vary across revisions and service updates, affecting compilation success and downstream API coverage. To mitigate this, we fix the model temperature to 0.0 and report results from a single fixed model version. Additionally, our API coverage metric counts only rustdoc-visible public APIs after two filtering steps; APIs exposed through re-exports, macros, or non-default features may not be captured, potentially over- or under-estimating the true coverage achievable (RQ1--RQ3).

\paragraph{Bug-finding scope.}
This evaluation measures fuzz target quality---compilability and API coverage---rather than vulnerability discovery, consistent with the evaluation methodology of RULF~\cite{rulf} and RPG~\cite{rpg}. The benchmark crates are mature, widely-used libraries in which latent bugs are unlikely to manifest within a 3-hour AFL++ session per target. A crash-oriented evaluation would require extended fuzzing budgets, bug-injection benchmarks, or evaluation on known-vulnerable library versions; we consider this complementary future work.

%% file: chapters/related_work.tex
\section{Related Work}

Despite Rust’s strong safety guarantees, prior work has shown that Rust libraries may still contain bugs and security vulnerabilities \cite{CloserLookSecurityRisks, RustForLinux, IsRustUsedSafely, UnderstandingMemoryAnd, Xu2020MemorySafetyCC}.
Automated techniques for systematically testing such libraries are therefore needed, particularly for crates that expose complex generic APIs.

\subsection{Static Analysis}

Static analysis detects defects without execution, using foundations such as abstract interpretation, data-flow analysis, model checking, and symbolic execution~\cite{abstract-interpretation,dataflow,static-analysis-survey,model-checking,symbolic-execution}.
For Rust, prior work formally models ownership and borrowing semantics~\cite{rustbelt,stacked-borrows,unsafe-rust} and builds practical analyzers on Rust's Mid-level Intermediate Representation (MIR) and LLVM IR to detect undefined behavior and unsafe misuse~\cite{mir-analysis,llvm-rust,miri,borrow-checking,polonius}.
However, generics, trait bounds, and associated types significantly complicate static reasoning over Rust library APIs~\cite{api-misuse-survey,typestate}.

\subsection{Fuzz Target Generation}

Fuzz target generation focuses on constructing program entry points that enable effective fuzzing of libraries.
Unlike application fuzzing, library fuzzing requires synthesizing valid API call sequences and inputs that satisfy type and semantic constraints \cite{libfuzzer}.

Early approaches rely on manually written fuzz targets, which are labor-intensive and difficult to scale. Several automated approaches instead generate fuzz targets by analyzing API signatures or extracting call sequences from documentation and examples \cite{autoharness,doc-fuzzing,api-mining}.
Other approaches construct fuzz targets by reusing values across API calls to improve coverage \cite{value-reuse,pool-based}.

For Rust, RULF generates fuzz targets by leveraging Rust’s type system and randomized API invocation strategies, enabling coverage-guided fuzzing of Rust libraries \cite{rulf}. Its Rust-specific fuzz target generation strategy makes it a strong baseline for library fuzzing, but our evaluation shows that it can still suffer from low coverage on highly generic crates and from generation failures on some targets.
RPG further introduces a pool-based fuzz target generation method that incrementally constructs fuzz targets by reusing previously generated values, improving public API coverage \cite{rpg}. Compared with RULF, RPG offers stronger value reuse and generic support in practice, and it even slightly outperforms \tool{} on \texttt{idna} in our evaluation; however, our experiments also show crate-level failures on several harder libraries.
deepSURF~\cite{deepsurf} is a concurrent LLM-augmented Rust fuzzing tool that employs a custom compiler plugin (\texttt{rustc\_surf}) to perform static analysis and identify unsafe-reaching APIs (URAPIs), then uses an LLM to generate fuzz targets targeting these APIs. deepSURF focuses specifically on reaching unsafe code blocks for memory-safety vulnerability detection, and does not aim to maximize general API coverage; it produces no fuzz targets for crates in which no unsafe-reaching APIs are detected.
Recent work also explores dependency-aware and structure-guided fuzz target synthesis to ensure type correctness and deeper API exploration \cite{api-fuzzing,structure-aware-fuzzing,graph-fuzzing}. In contrast to RULF's syntactic matching, RPG's pool-based reuse, and deepSURF's URAPI-targeted LLM fuzz targets, \tool{} combines explicit recursive generics-aware type matching with topology-guided dependency traversal and LLM-based concretization, targeting both compilability and broad API coverage.

\subsection{Fuzzing}

Fuzzing is a widely used dynamic testing technique that discovers bugs by executing programs with mutated inputs.
Coverage-guided fuzzers such as AFL and AFL++ have demonstrated significant success in uncovering real-world vulnerabilities across diverse software systems \cite{afl,aflplusplus}.
LibFuzzer enables in-process fuzzing and is commonly used for library testing \cite{libfuzzer}.

Numerous extensions improve fuzzing effectiveness by incorporating additional program analysis.
Angora introduces gradient-guided input mutation to solve complex branch conditions \cite{angora}.
Driller and QSYM combine fuzzing with symbolic execution to overcome path constraints and improve coverage \cite{driller,qsym}.
Other hybrid approaches integrate taint analysis or concolic execution to guide mutation strategies \cite{taint-fuzzing,hybrid-fuzzing}.

Structure-aware fuzzing techniques aim to preserve input validity by leveraging grammars or input models \cite{grammar-fuzzing,superion}.
However, these approaches primarily target input formats rather than API-level program structure.
For Rust libraries, the effectiveness of fuzzing depends on the quality of fuzz targets.
Poorly constructed targets may fail to compile, panic prematurely, or exercise only shallow API paths.

%% file: chapters/conclusion.tex
\section{Conclusion}

We presented \tool{}, a generics-aware fuzz target generation framework for Rust libraries. \tool{} extracts structured API information from Rust documentation, constructs an API dependency graph via a recursive type matching algorithm that explicitly handles generic parameters and trait bounds, and synthesizes compilable fuzz targets from dependency-guided API sequences using a large language model.

Evaluation on 13 real-world Rust libraries shows that \tool{} achieves a macro-average API coverage of 80.75\% and a macro-average compilation success rate of 96.19\%. On the eight crates where GRAFT, RULF, and RPG all produce runnable fuzz targets, \tool{} achieves 4.76$\times$ and 2.43$\times$ the average API coverage of RULF and RPG, respectively, while being the only tool to generate fuzz targets for all evaluated crates. On the eight crates with unsafe-reaching APIs, \tool{} outperforms deepSURF---a concurrent LLM-augmented tool designed for memory-safety vulnerability detection---by 1.41$\times$ in average API coverage. Ablation studies confirm that each of the three core components---generics-aware type matching, the API dependency graph, and topology-guided traversal---contributes independently to coverage quality.

One natural extension is cross-crate API interaction, where type-flow edges span crate boundaries. Runtime feedback from fuzzing runs could further guide API sequence selection, and the dependency-graph approach may generalize to detecting semantic bugs beyond memory-safety violations.